# Arbitrary order exceptional point induced by photonic spin-orbit interaction in coupled resonators

Shubo Wang[1,*], Bo Hou[2,3], Weixin Lu[2], Yuntian Chen[4], Z. Q. Zhang[3], and C. T. Chan[3,*]

[1] *Department of Physics, City University of Hong Kong, Hong Kong, China*
[2] *School of Physical Science and Technology & Collaborative Innovation Center of Suzhou Nano Science and Technology, Soochow University, Suzhou, China*
[3] *Department of Physics, The Hong Kong University of Science and Technology, Hong Kong, China*
[4] *School of Optical and Electronic Information, Huazhong University of Science and Technology, Wuhan, China*

*Correspondence should be addressed to:
Shubo Wang (email: shubwang@cityu.edu.hk) or C. T. Chan (email: phchan@ust.hk)

**Abstract**

Many novel properties of non-Hermitian systems are found at or near the exceptional points—branch points of complex energy surfaces at which eigenvalues and eigenvectors coalesce. In particular, higher-order exceptional points can result in optical structures that are ultrasensitive to external perturbations. Here we show that an arbitrary order exceptional point can be achieved in a simple system consisting of identical resonators placed near a waveguide. Unidirectional coupling between any two chiral dipolar states of the resonators mediated by the waveguide mode leads to the exceptional point, which is protected by the transverse spin-momentum locking of the guided wave and is independent of the positions of the resonators. Various analytic response functions of the resonators at the exceptional points are experimentally manifested in the microwave regime. The enhancement of sensitivity to external perturbations near the exceptional point is also numerically and analytically demonstrated.

**Introduction**

If the Hamiltonian of a system takes the form of a Hermitian matrix, it has only real eigenvalues accompanied by a set of orthogonal eigenvectors. If the Hamiltonian is non-Hermitian, its eigenvalues are complex in general. By tuning the amount of non-Hermiticity, one can make two eigenvalues coalesce at one point, which is called an exceptional point (EP)[1–3]. In contrast to the degeneracy in a Hermitian system (a diabolic point), an EP is a branch point in the complex energy surface where the Hamiltonian matrix becomes defective. EPs were found in non-Hermitian systems possessing parity-time ($PT$) symmetry, where the simultaneous increase of gain and loss can induce symmetry breaking, characterised by a change of frequency spectrum from being purely real to being complex[4,5]. EPs can also exist in passive systems without $PT$ symmetry as a result of the unbalanced competition between loss contrast and coupling strength of two states[6]. The phase transition associated with EPs leads to numerous counterintuitive phenomena such as loss (pump)-induced revival (depressing) of lasing[6,7], optical isolation[8], unidirectional transport of light[4,9–11], topological energy transfer[12] and others[13–21]. Recently, higher-order EPs have also attracted a lot of attention[22–27]. A higher-order EP is a result of coalescence of multiple eigenvalues[22], which normally requires a delicate variation of parameters in a larger parameter space and is much more difficult to achieve[25,28].

For a simple two-level non-Hermitian system with a Hamiltonian of the form $H = \begin{bmatrix} a & c \\ d & b \end{bmatrix}$, an EP can be obtained under the condition of $(a-b)^2 + 4cd = 0$. This can be achieved when $\text{Re}(a) = \text{Re}(b)$, $a-b$ is purely imaginary and $c = d^*$ [4,6–9] or when $a-b$ is real and $c = d$ is purely imaginary[28–31]. In the former case, the non-Hermiticity comes from the asymmetric loss/gain of the two identical states, whereas in the latter case the non-Hermiticity appears in the coupling of two states mediated by some open channels. However, EPs can also be achieved when $a = b$ and either $c = 0$ or $d = 0$, corresponding to a unidirectional coupling between two states. EPs of this kind can be achieved in systems with structural asymmetry[32,33].

In this paper, we propose a different scenario to achieve unidirectional couplings. Our systems consist of a set of identical dipolar resonators sitting on a strip waveguide. The unidirectional coupling between any two resonators arises from the evanescent couplings between the chiral dipolar modes of the resonators and the waveguide mode. The transverse spin-momentum locking of the guided wave excludes the backward couplings among the resonators. EPs achieved in this manner can be of arbitrary order, depending only on the number of resonators. Although the positions of the resonators do not affect the order of an EP, they do determine the spectral response functions of the resonators. We have studied such systems theoretically using both coupled mode theory (CMT)[34,35] and full-wave simulations. An experimental demonstration of the theory is also carried out in a system consisting of multiple high-dielectric spheres on a strip waveguide in the microwave regime.

**Results**

**Exceptional points in a model system.** We first consider a model system consisting of two identical spheres located near a strip waveguide that forms a circle, as shown in Fig. 1a. The purpose of studying this model system is twofold. It not only proves that the model system in the next subsection, where EP appears naturally, is just a limiting case of the model system in Fig. 1a, but also uncovers the critical role of unidirectional coupling in achieving arbitrary order EPs. In cylindrical coordinates $(r,\theta,z)$, the waveguide's geometry is defined by $R \leq r \leq R+t$ and $-w/2 \leq z \leq w/2$, with $t$ being the waveguide's thickness and $w$ being the waveguide's width. The waveguide has a rectangular cross section and is made of gold. The positions of the two spheres are denoted by two azimuthal angles $\theta_1 = 3\pi/2$ and $\theta_2 = \pi/2$. The permittivities of the spheres are described by the Drude model, which gives an electric dipole resonance at about 300 THz.

The strip waveguide shown in Fig. 1a supports guided plasmon modes. The dispersion relation of the lowest-order mode is shown in Fig. 1b by the solid blue line, where $n_{\text{eff}} = k_{\text{plas}}/k_0$ is the effective mode index, with $k_{\text{plas}}$ being the mode propagation constant and $k_0$ being the vacuum wave number. The corresponding mode electric field is shown in the inset of Fig. 1b, where the colour denotes the field magnitude and the arrows indicate field directions. We note that the electric field is well confined around the waveguide, facilitating long-range coupling between any two dipole resonators through the guided plasmon[36]. Because the plasmon field (i.e. evanescent electromagnetic field) in Fig. 1a has both transverse ($E_r$) and longitudinal ($E_\theta$) electric field components, it carries intrinsic spin angular momentum (SAM) in the $z$ direction, i.e., the plasmon field is in general elliptically polarised[37]. The SAM is determined by the material and geometrical properties of the waveguide. It has been shown that such a propagating plasmon can be mapped into a quantum spin Hall system with the property of spin-momentum locking (plasmon carrying opposite SAM propagating in opposite directions)[38,39].

Each sphere in Fig. 1a supports two degenerate electric dipole modes along $\hat{x}$ and $\hat{y}$ directions, respectively. A superposition of the two modes gives rise to a pair of chiral dipole modes $\mathbf{p} = p\hat{e}_\pm$, where $\hat{e}_\pm = (\hat{x} \pm i\hat{y})/\sqrt{2}$ is the circular basis of the spheres[40–43]. Such chiral modes should not be confused with the eigenmodes of a sphere made of chiral materials, which has intrinsic chirality[44]. Below, we will focus on the interaction between such chiral dipole modes mediated by the waveguide mode. In the presence of the strip waveguide, the resonant chiral dipole field asymmetrically couples to the propagating plasmon of the waveguide[45]. We carefully designed the waveguide (tuning material parameters and cross-sectional geometry) so that the SAM of the chiral dipole field matches the SAM of the guided plasmon. Because of the locking between the SAM and the linear orbital momentum (**k**), the excited plasmon propagates in one direction only. Such a process can be regarded as the conversion of light's SAM to linear orbital momentum (i.e., photonic spin-orbit interaction[45]), which can produce many interesting applications/phenomena such as chiral photonic waveguides[46] and lateral optical forces[44,47,48]. The resulting coupling between the two spheres is unidirectional, i.e., either clockwise or counter-clockwise, as shown in Fig. 1a, depending on the SAM of the incident light. The system can be described by an effective Hamiltonian

$$H = \begin{bmatrix} \omega_0 - \frac{i}{2}(\gamma_1 + \gamma_c) & \kappa_{12} \\ \kappa_{21} & \omega_0 - \frac{i}{2}(\gamma_2 + \gamma_c) \end{bmatrix}. \quad (1)$$

Here, $\omega_0$ is the dipole resonance frequency of the isolated spheres, $\gamma_{1,2}$ and $\gamma_c$ denote the material loss and radiation loss, respectively, of the spheres and $\kappa_{21}$ ($\kappa_{12}$) denotes the coupling parameter from Sphere 1 (2) to Sphere 2 (1), as shown in Fig. 1a for the case under left-handed circular polarisation (LCP) excitation. Here, the coupling parameters have complex values in general, which are in contrast to those systems where the mode couplings are realised through near-field hopping[6,49] and are usually real-valued. The corresponding eigenfrequencies of Eq. (1) are $\omega_{\pm} = \omega_0 - i\Gamma/2 \pm \sqrt{\kappa_{12}\kappa_{21} - (\Delta\gamma/4)^2}$, where $\Gamma = (\gamma_1 + \gamma_2 + 2\gamma_c)/2$ and $\Delta\gamma = \gamma_2 - \gamma_1$ denotes the loss contrast of the two spheres. Using CMT, the amplitudes of the electric chiral dipole moments can be determined as (see Methods)

$$p_1 = \left| \frac{2\sqrt{\gamma_c} A_{in} \left[ -2i\kappa_{12} + \Gamma - 2i(\omega - \omega_0) \right]}{4\kappa_{12}\kappa_{21} + \left[ \Gamma - 2i(\omega - \omega_0) \right]^2} \right|, \quad p_2 = \left| \frac{2\sqrt{\gamma_c} A_{in} \left[ -2i\kappa_{21} + \Gamma - 2i(\omega - \omega_0) \right]}{4\kappa_{12}\kappa_{21} + \left[ \Gamma - 2i(\omega - \omega_0) \right]^2} \right|, \quad (2)$$

where $A_{in}$ is the incident field amplitude.

In general, the Hamiltonian in Eq. (1) is non-Hermitian even if the system has no loss. The non-Hermiticity here originates from the unidirectional coupling which leads to $\kappa_{12} \neq (\kappa_{21})^*$. We note that this does not imply a broken reciprocity since the Hamiltonian only describes a subsystem corresponding to the coupling of one chiral mode. The other chiral mode (with opposite chirality) couples via the plasmon propagating in the opposite direction. The two scenarios have equal coupling strength and therefore the system is reciprocal[40,50]. The value of $\omega_\pm$ can be tuned by varying the product $\kappa_{12}\kappa_{21}$, which can be expressed as $\kappa_{12}\kappa_{21} = \left( -i\kappa_{12}^0 e^{i\phi_{12}} \right)\left( -i\kappa_{21}^0 e^{i\phi_{21}} \right) = -\kappa_{12}^0 \kappa_{21}^0 e^{i(\phi_{12}+\phi_{21})}$.[51] Here, $\kappa_{12}^0$ and $\kappa_{21}^0$ are the positive real-valued coupling strengths and $\phi_{12} + \phi_{21}$ accounts for the total dynamic phase accumulated by the guided plasmon. An EP can be achieved when $\kappa_{12}\kappa_{21} = (\Delta\gamma/4)^2$. Since the loss contrast $\Delta\gamma$ has a real value, this requires $\kappa_{12}\kappa_{21}$ to be real and positive, therefore, $\phi_{12} + \phi_{21} = (2m+1)\pi, m = 0, 1, ...$, which corresponds to destructive interference of the plasmon. Such a condition is easily fulfilled by tuning the radius $R$ of the waveguide. Figure 1c shows the plasmon amplitude when the circular waveguide in Fig. 1a is excited by a point source. The peaks correspond to constructive interferences similar to the Mie resonances associated with dielectric spheres/disks. The dips marked by the red arrows correspond to destructive interferences, where $\kappa_{12}\kappa_{21} = \kappa_{12}^0 \kappa_{21}^0$. The value of the product $\kappa_{12}^0 \kappa_{21}^0$ can be controlled by tuning the loss of the waveguide. Therefore, it is possible to achieve an EP when $\kappa_{12}\kappa_{21} = \kappa_{12}^0 \kappa_{21}^0 = (\Delta\gamma/4)^2$. To demonstrate such a possibility, we did full-wave simulations of the system shown in Fig. 1a with $R = 200$ nm under the incidence of a plane wave of LCP, in which case the left-handed chiral dipole modes were excited and coupled with each other. The electric dipole moments of the spheres were calculated and fitted to the CMT expressions in Eq. (2), from which we obtained the values of $\gamma_{1,2}$, $\gamma_c$ and $\kappa_{12}, \kappa_{21}$. In the simulations, we only introduced loss into the lower-half waveguide ($\theta \in [\pi/2, 3\pi/2]$) shown in Fig. 1a to suppress $\kappa_{12}\kappa_{21}$ (see Methods for details). When the introduced waveguide loss $\gamma_{wg}$ was increased, an EP emerged at $\kappa_{12}\kappa_{21} = (\Delta\gamma/4)^2$, where the real parts of the eigenfrequencies merge and their imaginary parts bifurcate. Note that $\gamma_{wg}$ contributes little to the phase of $\kappa_{12}\kappa_{21}$ (see Supplementary Figure 1). Figure 1d shows such a scenario for different values of $\Delta\gamma$. As $\Delta\gamma$ is reduced, the EP moves to the right with an enlarged splitting of $\text{Re}(\omega_\pm)$ and reduced splitting of $\text{Im}(\omega_\pm)$. In the limit of $\Delta\gamma \to 0$, i.e., approaching the limit of identical spheres, the EP will be reached when $\kappa_{12} \to 0$ ($\gamma_{wg}$ is sufficiently large so that the coupling from $p_2$ to $p_1$ is completely suppressed). In this case, the system Hamiltonian is reduced to a Jordan block form

$$H = \begin{bmatrix} \tilde{\omega}_0 & 0 \\ \kappa_{21} & \tilde{\omega}_0 \end{bmatrix}, \text{ where } \tilde{\omega}_0 = \omega_0 - i\Gamma/2, \text{ and Eq. (2) reduces to}$$

$$p_1 = \left| \frac{i\sqrt{\gamma_c} A_{in}}{\omega - \tilde{\omega}_0} \right|, \quad p_2 = \left| \frac{i\sqrt{\gamma_c} A_{in}}{\omega - \tilde{\omega}_0} + \frac{i\sqrt{\gamma_c} A_{in} \kappa_{21}}{(\omega - \tilde{\omega}_0)^2} \right|. \tag{3}$$

Different from Eq. (2), Eq. (3) exhibits a single pole but of different orders for $p_1$ and $p_2$. The presence of a second-order pole in $p_2$ is a signature of EPs[51,52]. It is noted that the unidirectional coupling renders $p_1$ unperturbed, whereas it contributes to $p_2$. The interference between the first-order and second-order terms in $p_2$ can give rise to an asymmetric spectrum different from the Lorentzian response in $p_1$[53]. Below, we will focus on EPs in the case of unidirectional coupling.

**Exceptional points achieved with unidirectional coupling.** At the EP of the case where $\Delta\gamma \rightarrow 0$, the coupling from $p_2$ to $p_1$ is suppressed by the lossy waveguide, and thus the model system shown in Fig. 1a can be simplified to that shown in Fig. 2a, where the two spheres are placed on a straight waveguide and are separated by a distance $d$. The guided plasmon propagates unidirectionally under spin-momentum locking and is absorbed at the end of the waveguide under an absorbing boundary condition. The absorption of the plasmon prevents reflection at the end of the waveguide, and the coupling between spheres is unidirectional (i.e. $\kappa_{12} = 0$). We carried out full-wave simulations for such a configuration. The simulation details and system parameters are given in the Method section. The electric dipole moments of the two spheres are shown by the blue ($p_1$) and red ($p_2$) symbols in Figs. 2b–k for $d = 2.0\lambda_{plas}$ (Figs. 2b, c), $d = 2.2\lambda_{plas}$ (Figs. 2d, e), $d = 2.4\lambda_{plas}$ (Figs. 2f, g), $d = 2.6\lambda_{plas}$ (Figs. 2h, i), and $d = 2.8\lambda_{plas}$ (Figs. 2j, k). Here, $\lambda_{plas} = 2\pi / k_{plas}$ is the wavelength of the guided plasmon. For $d = 3.0\lambda_{plas}$, the spectra of two dipole moments are identical to those of $d = 2.0\lambda_{plas}$, exhibiting a periodic pattern with $\lambda_{plas}$ being the period. We note that $p_1$ has a typical Lorentzian shape and remains unchanged when $d$ changes, whereas $p_2$ changes dramatically because of the interference effect under unidirectional coupling. The CMT results are also shown in Figs. 2b–k as solid lines, which match well with the simulation results. In Fig. 2c, $p_2$ shows a dip at the resonance frequency, which can be understood from Eq. (3). At $\omega = \omega_0$, Eq. (3) gives $p_2 = \left| 4\sqrt{\gamma_c} A_{in} \left( \Gamma / 2 - i\kappa_{21} \right) / \Gamma^2 \right|$. The variation of $d$ only changes the phase of $\kappa_{21}$. Figure 2l shows the evolution of the function $Z(d) = \Gamma / 2 - i\kappa_{21}(d)$ on the complex plane when $d$ changes from $d = 2.0\lambda_{plas}$ to $d = 2.8\lambda_{plas}$, corresponding to Figs. 2b–k. The evolution trajectory of $Z(d)$, which is denoted by the dashed circle in Fig. 2l, can be well described by using the expression $\kappa_{21}(d) = -i\kappa_{21}^0 \exp(ik_{plas}d)$. The arrow on the circle shows the evolution direction. The phase factor of $\kappa_{21}(d)$ results from the propagation of the plasmon from Sphere 1 to Sphere 2. The fitted values of $\kappa_{21}(d)$ obtained from Figs. 2b–k are marked by circular symbols in Fig. 2l. When $d$ is an integer multiple of $\lambda_{plas}$, $\kappa_{21}$ has no phase contribution from the guided plasmon and we have $\kappa_{21} = -i\kappa_{21}^0$ [51], and therefore $Z(d)$ has a minimum value of $\Gamma/2 - \kappa_{21}^0$, which explains the dip of $p_2$ in Fig. 2c. When $d$ is a half-integer multiple of $\lambda_{plas}$, $Z(d)$ has a maximum value of $\Gamma/2 + \kappa_{21}^0$, which explains the enhancement of $p_2$ in Fig. 2g. Note that we have ignored the absorption of the waveguide in Fig. 2a for simplicity, but the results would be qualitatively the same if absorption is also considered.

**Higher-order exceptional points**. The realisation of higher-order EPs is hotly sought after, not only for the interesting physics, which goes beyond the square-root singularity, but also for their application potential such as for ultrasensitive sensors[32,54–57]. In conventional non-Hermitian systems, the realisation of higher-order EPs is usually very difficult as it involves complicated design and fine tuning of many parameters[22,25,26]. We show that EPs of arbitrary order can be readily achieved in our system by introducing more identical resonators on the waveguide. For $N$ identical spheres on the waveguide with unidirectional couplings, as shown in Fig. 3a, all of the coupling parameters $\kappa_{ij}, i < j$ vanish so that the effective Hamiltonian can be expressed as

$$H = \begin{bmatrix} \tilde{\omega}_0 & 0 & \cdots & 0 \\ \kappa_{21} & \tilde{\omega}_0 & \cdots & 0 \\ \vdots & \vdots & \ddots & \vdots \\ \kappa_{N1} & \kappa_{N2} & \cdots & \tilde{\omega}_0 \end{bmatrix}. \tag{4}$$

The Hamiltonian matrix in Eq. (4) represents an EP of order $N$. It is defective, having a single eigenvalue $\tilde{\omega}_0 = \omega_0 - i\Gamma/2$ with algebraic multiplicity $N$, i.e., the $N$ eigenvalues coalesce. Thus, the order of the EP can be increased simply by putting more resonators on the waveguide. The coupling constant $\kappa_{ij}$ will only affect the forms of the spectral response functions of the resonators. Here, we consider an example with five equally spaced spheres with a separation $d = \lambda_{\text{plas}}$, as shown in Fig. 3a. Figure 3b shows the electric dipole moments of the five spheres under LCP excitation. The solid lines denote the results obtained from CMT with the Hamiltonian in Eq. (4). The symbols denote the results of full-wave simulations. Excellent agreement between the two is found. It is noted that only $p_1$ shows a Lorentzian shape, whereas the rest are suppressed because of interference. Therefore, at the resonance frequency, the dipole moment as a function of the resonator's position has a monotonic trend, as shown in Fig.3c, where each point denotes the dipole moment magnitude of the corresponding sphere. Under LCP excitation, the dipole moment gradually decreases from left to right, whereas this trend is reversed under right-handed circular polarisation (RCP). Such a phenomenon can be understood based on CMT. Because the period of the array is $d = \lambda_{\text{plas}}$, all of the coupling parameters in Eq. (4) have zero phase, i.e., $\kappa_{ij} = \kappa_{21} = -i\kappa_{21}^0, \forall i > j$. Thus, at $\omega = \omega_0$, one can easily derive the expressions for the resonant dipole moments using CMT (Supplementary Note 1) as $p_i = \left| 2\sqrt{\gamma_c}(A_{\text{in}}/\Gamma)\left[(\Gamma - 2\kappa_{21}^0)/\Gamma\right]^{i-1} \right|$. Because both $\kappa_{21}^0$ and $\Gamma$ take positive real values and $\left|(\Gamma - 2\kappa_{21}^0)/\Gamma\right| < 1$ according to the results in Fig. 2l, we have $p_i < p_{i-1}$, which indicates a monotonic decrease of the dipole moments and the decrease is reversed for the RCP excitation by mirror symmetry. Such a decrease can also be understood as a kind of cascade effect under destructive interferences between the spheres, where the coupling field from $p_i$ suppresses the field of the spheres on its right side. It is also possible to have constructive interference for the spheres, such that the sphere at the end has much larger emission power than an isolated sphere (see Supplementary Figure 2). The mechanism revealed here enables spin-controlled dipole array emission and energy accumulation in the lateral direction.

Higher-order EPs are characterised by an enhanced sensitivity to external perturbations. In the case of small perturbations, the eigenfrequency splitting induced near an EP of order $N$ are of the order $\sqrt[N]{\epsilon}$, where $\epsilon$ denotes the strength of the small perturbation. Because the order of EP in our system is directly determined by the number of spheres $N$, the corresponding sensitivity can be easily controlled. To simplify the numerical calculation, we demonstrate the higher-order sensitivity enhancement in our system by considering a two-dimensional (2D) version of our model system, i.e., five parallel cylinders on a 2D metal waveguide, as shown in Fig. 4a. The cylinders are equally spaced with a distance of $d = 5\lambda_{plas}$. The waveguide is truncated on the right side with a distance of $5\lambda_{plas}$ between its edge and the last cylinder. An absorption boundary condition is applied on the left edge. Because the waveguide truncation on the right edge acts as a mirror, a propagating LCP plasmon from left to right will be totally reflected by the truncated edge. Because the reflected plasmon is of RCP, it will also excite the RCP modes of the five cylinders. Thus, the system involves 10 unidirectionally coupled modes with five LCP modes travelling to the right and five RCP modes travelling to the left and is at an EP of order 10.

To demonstrate the enhanced sensitivity of order $\sqrt[10]{\epsilon}$, perturbation is introduced by inserting a small parallel cylinder with an elliptical cross section between the first cylinder and the left edge. The strength of perturbation $\epsilon$ is proportional to the reflection caused by the small elliptical cylinder, which can be controlled by the amount of loss in the small cylinder (see Methods). Figure 4b, with results calculated numerically using COMSOL, shows the spectra of total electric dipole moment strength of the system at different values of $\epsilon$. We notice that as $\epsilon$ increases, more resonance peaks appear in the spectrum, and the splitting between peaks increases as well. When $\epsilon$ is large enough, a total of 10 peaks are conspicuous, corresponding to 10 eigenvalues, as expected. In Fig. 4c, we plot the splitting of the eigenvalues $\Delta\omega_n = |\omega_n - \omega_0|$ as a function of $\epsilon$ for the first and the last peaks shown in Fig. 4b in log–log scales. For small $\epsilon$ values, the results show a straight line with a slope close to 1/10, i.e., $\sqrt[10]{\epsilon}$ behaviour for an EP of order 10.

The above system can be described by the following Hamiltonian:

$$H = \begin{bmatrix} H^{\mathrm{L}} & \kappa^{\mathrm{LR}} \\ \kappa^{\mathrm{RL}} & H^{\mathrm{R}} \end{bmatrix}, \tag{5}$$

where the two $N \times N$ block matrices $H^{\mathrm{L}}$ and $H^{\mathrm{R}}$ denote the unidirectionally coupled $N$ LCP modes and $N$ RCP modes, respectively. They both have the form shown in Eq. (4), with $N = 5$ and $\kappa_{ij} = -i\kappa_{21}^0$. The block matrix $\kappa^{\mathrm{RL}}$ denotes the one-way conversion from LCP modes to RCP modes occurring at the right edge of the waveguide and $\kappa^{\mathrm{LR}}$ denotes the excitation of right-going LCP modes at the perturbation cylinder on the left side of the waveguide caused by reflection. These two block matrices have the forms $\kappa_{ij}^{\mathrm{RL}} = -i\kappa_{21}^0$ and $\kappa_{ij}^{\mathrm{LR}} = -i\epsilon\kappa_{21}^0$, respectively. Note that the corresponding eigenvectors have the form of $\left[a_1^{\mathrm{L}}, a_2^{\mathrm{L}}, \ldots, a_5^{\mathrm{L}}, a_5^{\mathrm{R}}, a_4^{\mathrm{R}}, \ldots, a_1^{\mathrm{R}}\right]^{\mathrm{T}}$. Figure 4d shows the real parts of the normalised eigenvalues $\omega_n / \kappa_{21}^0$ as a function of perturbation strength $\epsilon$, which exhibits the eigenvalue splitting phenomenon shown in Fig. 4b. Note that $\omega_5$ and $\omega_6$ have identical real parts. In the numerical simulations, all 10 eigenvalues can be observed because the spacing between cylinders is not exactly an integer multiple of plasmon wavelength. In Fig. 4e, we plot the splitting of $\Delta\omega_n / \kappa_{21}^0$ as a function of $\epsilon$ in log–log scales, where only $n = 1, 2, 3, 4$ is considered because of the symmetric splitting of the eigenvalues. Enhanced sensitivity of order $\sqrt[10]{\epsilon}$ can be clearly seen in the small $\epsilon$ region, where all of the splitting follows a trend with a slope of 1/10. When the perturbation is large, only $\Delta\omega_1 / \kappa_{21}^0$ maintains a 1/10 slope, as shown by the zoom-in insert in Fig. 4e, in accordance with the numerical results in Fig. 4c. The trends of the other eigenvalue splitting deviate from the 1/10 slope because of the contributions of high-order perturbation terms.

**Experimental demonstration at microwave frequencies.** The physics discussed above also apply to the microwave regime. We note that the magnetic field of light can also carry spin angular momentum. Consider a dielectric sphere which supports two orthogonal magnetic dipole modes along $\hat{x}$ and $\hat{y}$ directions, respectively. A superposition of the two modes gives rise to a pair of chiral magnetic dipole modes $\mathbf{m} = m\hat{e}_{\pm}$, one of which is shown by the inset in the lower right corner of Fig. 5a. A simple dielectric waveguide can support guided wave with evanescent magnetic field carrying spin. The combination of the dielectric spheres and dielectric waveguide then corresponds to the magnetic version of the plasmonic mode system and can be easily realized at microwave frequencies. We carried out an experimental demonstration of the properties associated with the EPs in our system using the setup shown in Fig. 5a. We used high-dielectric spheres as the resonators, which have a magnetic dipole resonance near 9 GHz. The waveguide was also made of a dielectric material, and its fundamental mode served to guide the wave in the considered frequency region. The two ends of the waveguide were inserted into absorbing materials to suppress the reflected waves. The chiral magnetic dipole mode was excited using a circularly polarised microwave incident along the $z$ direction. To measure the amplitude of the magnetic dipole moments, a probe was used to detect the magnetic field above the spheres (see Fig. 5a).

Figure 5b shows the magnitude of the magnetic field near the left and right spheres, respectively, under LCP excitation. The results of RCP excitation have similar features and are shown in Supplementary Figure 3. The symbols here denote the experimental results and the solid lines denote the analytical fitting results based on Eq. (3). The measurements were done at different coupling distances $d$. We note that as the coupling distance changes from $d = 50$ mm to $d = 60$ mm, the left dipole resonance remains unchanged, whereas the other dipole undergoes dramatic variations, in accordance with the numerical results shown in Figs. 2b–k. The position of the peak for the left sphere (marked by a dashed line) is a constant. The interference feature of the right sphere gradually changes as $d$ increases. This can be seen from the positions of the maximum and minimum values near the resonance frequency $\omega_0$, shown in Fig. 5c, which undergoes a red-shift when $d$ increases. Note that there is a switch of the maximum at $d = 2\lambda_{wg} = 55$ mm, with $\lambda_{wg}$ being the wavelength of the guide wave. The experimental results match the analytical CMT results very well, as shown in Fig. 5c. The asymmetry of the spectra, as well as the evolution of the maximum/minimum for the right sphere, can be explained by Eq. (2) with $\kappa_{12} = 0$. The numerator of $p_2$ has a factor of $\Gamma/2 - i\kappa_{21} - i\Delta\omega$, where $\Delta\omega = \omega - \omega_0$. Because $\kappa_{21}$ is a complex number, it is obvious that $p_2$ is asymmetric with respect to $\Delta\omega = 0$ except for the special case of $\text{Re}(\kappa_{21}) = 0$. This can be seen from Fig. 5d, where the circular symbols denote the values of function $Z(d) = \Gamma/2 - i\kappa_{21}$ on the complex plane, obtained from the fitting results in Fig. 5b. For any case with $\text{Re}(\kappa_{21}) \neq 0$, the values of $|\Gamma/2 - i\kappa_{21} + i\Delta\omega|$ and $|\Gamma/2 - i\kappa_{21} - i\Delta\omega|$ (denoted by the length of the red arrows) are not equal, which indicates that $p_2$ has a shape that is asymmetric with respect to centre frequency $\omega_0$. It is the looping of the complex coupling parameter $\kappa_{21}$ that accounts for the shift of the minimum/maximum in Fig. 5b and Fig. 5c. We note that similar phenomena also exist in the plasmonic model system shown in Fig. 2a except that the spectra asymmetry of $p_2$ is weak due to a smaller coupling strength $\kappa_{21}^0$ compared with $\Gamma/2$ (corresponding to a smaller circle in Fig. 2l). The experiment confirmed the mechanism of unidirectional coupling between identical resonators which can be described by the Hamiltonian in Eq. (4) (where an EP naturally arises). It also demonstrated the interference effect between the first-order and second-order poles in Eq. (3). We emphasize that the second-order pole emerges in the Green's function as a direct result of the coalescence of two eigenvalues at the EP[52,58]. The interference effect at the EP is different from the ordinary interference of two resonators without an EP. In the latter case, the spectra of both resonators should be asymmetric in general and depend on the coupling. Here, as shown in Fig. 5b, the spectrum of Sphere 1 is approximately symmetric and does not depend on the value of coupling distance $d$, while the spectrum of Sphere 2 is asymmetric and depends on the value of $d$.

We also performed measurements for the higher-order EP, which manifested spin-controlled dipole array emission. Because of the finite length of the waveguide, we put four spheres on the waveguide with equal spacings of $\lambda_{wg}$ and measured the resonant magnetic field near each sphere under LCP and RCP excitations. The results are shown by solid circles in Fig. 6. In accordance with the prediction in Fig. 3c, the magnetic field has a monotonic trend determined by the SAM of the incident microwave wave. The slight asymmetry between LCP and RCP results is owing to the imperfection of the microwave source wave front. We note that the field dependence in Fig. 3c and Fig. 6 are direct results of the interference effect in presence of higher-order poles at the EPs. We also note that, different from the dipole moment, the evanescent waves coupled from the spheres to the waveguide interference constructively since $d = \lambda_{plas}$, which leads to a growing field magnitude of the waveguide (Supplementary Figure 4).

**Discussion**

For a physical system described by a Hamiltonian involving many physical parameters, it typically requires tedious tuning of multiple parameters to achieve a higher-order EP. Worse still, in a real system, the parameters may be correlated, making their tuning even more complicated. This issue renders a higher-order EP rather difficult to achieve. One feasible way is to have a physical mechanism that guarantees all of the parameters fulfil the condition of the EPs. In our system, the mechanism is the unidirectional coupling of identical resonators induced by the locking between the light SAM and linear orbital momentum. With such a mechanism, the phenomenon associated with arbitrary order EPs can be realised simply by expanding the system to include more spheres. The higher-order EPs demonstrated here should have potential applications in sensing. In such applications, resonators with high quality factors are needed to detect small perturbations, although it is not necessary for a proof-of-concept demonstration of the EPs here. In the model system with unidirectional coupling between identical spheres, we have assumed a lossless plasmonic waveguide. When loss is taken into account, the coupling strength between spheres will be reduced, and the coupling parameters in the Hamiltonian will have different magnitudes in general. However, the Hamiltonian will remain in the same form (i.e. a triangular matrix), and the discussed physics about EPs still hold. In this case, one has to assemble the spheres closely to obtain strong EP phenomena.

In summary, we have demonstrated that the spin-orbit interaction of light can induce unidirectional coupling between chiral dipole modes of identical spherical resonators through a waveguide if the chiral modes are selectively excited using circularly polarised light. The unidirectional coupling leads to arbitrary order EPs depending on the number of resonators on the waveguide. Various analytic response functions of the resonators at the exceptional points have been experimentally manifested. The discussed mechanism also provides an easy way to control the excitation of the coupled resonance modes using the SAM of light. The proposed configuration can serve as a convenient platform for studying light–matter interactions at EPs and can be integrated into a photonic circuit for on-chip applications.

**Methods**

**Coupled mode theory for two dipole resonators.** The rate equations for the two chiral dipole modes can be written as[34,35]

$$\frac{da_1}{dt} = -i\omega_0 a_1 - \frac{\gamma_1 + \gamma_c}{2} a_1 - i\kappa_{12} a_2 - \sqrt{\gamma_c} a_{in}, \tag{6}$$

$$\frac{da_2}{dt} = -i\omega_0 a_2 - \frac{\gamma_2 + \gamma_c}{2} a_2 - i\kappa_{21} a_1 - \sqrt{\gamma_c} a_{in}, \tag{7}$$

where $a_i$ is the mode amplitude. The above equations can be re-written as

$$\frac{d\Lambda}{dt} = -iH\Lambda - \sqrt{\gamma_c} \Lambda_{in} \tag{8}$$

where

$$H = \begin{bmatrix} \omega_0 - \frac{i}{2}(\gamma_1 + \gamma_c) & \kappa_{12} \\ \kappa_{21} & \omega_0 - \frac{i}{2}(\gamma_2 + \gamma_c) \end{bmatrix}, \Lambda = \begin{bmatrix} a_1 \\ a_2 \end{bmatrix}, \Lambda_{in} = \begin{bmatrix} a_{in} \\ a_{in} \end{bmatrix} \tag{9}$$

Under the excitation of a harmonic wave with $a_{in} = A_{in} e^{-i\omega t}$, the steady-state solutions for $a_i = A_i e^{-i\omega t}$ obtained from Eqs. (6) and (7) are

$$A_1 = -\frac{2\sqrt{\gamma_c} A_{in}\left[-2i\kappa_{12} + \Gamma - 2i(\omega - \omega_0)\right]}{4\kappa_{12}\kappa_{21} + \left[\Gamma - 2i(\omega - \omega_0)\right]^2}, A_2 = -\frac{2\sqrt{\gamma_c} A_{in}\left[-2i\kappa_{21} + \Gamma - 2i(\omega - \omega_0)\right]}{4\kappa_{12}\kappa_{21} + \left[\Gamma - 2i(\omega - \omega_0)\right]^2}, \tag{10}$$

where $\Gamma = (2\gamma_c + \gamma_1 + \gamma_2)/2$. The magnitude of the induced dipole moments for each sphere are $p_1 = |A_1|$ and $p_2 = |A_2|$. In the case of unidirectional coupling, e.g. $\kappa_{12} = 0$ and $\kappa_{21} \neq 0$, we obtain Eq. (3).

**Numerical simulation.** For the plasmonic model systems shown in Fig. 1a and Fig. 2a, the spheres have diameters of 10 nm. The distance between the spheres and the waveguide is $h_p = 17$ nm. The permittivities of the spheres are described by the Drude model $\varepsilon_i = 1 - \omega_p^2/(\omega^2 + i\omega\gamma_i)$, where $\omega_p = 3.3 \times 10^{15}$ rad s$^{-1}$ and the sub-script '$i$' denotes the $i$th sphere. For the system in Fig. 1a, we set $\gamma_1 = 1.0 \times 10^{13}$ rad s$^{-1}$ and $\gamma_2 = \gamma_1 + \Delta\gamma$ with $\Delta\gamma$ being the loss difference in Fig. 1d. For the system in Fig. 2a, we set $\gamma_1 = \gamma_2 = 1.0 \times 10^{13}$ rad s$^{-1}$. The waveguide in Fig. 1a has a cross section of $w \times t = 20$ nm $\times 5$ nm and forms a circle of radius $R = 200$ nm. We set $\varepsilon_{wg} = -50$ for the upper half of the waveguide and $\varepsilon_{wg} = -50 + 50i\gamma_{wg}(1 - 2|\theta - \pi|/\pi), \theta \in [\pi/2, 3\pi/2]$ for the lower half. The waveguide in Fig. 2a has the same cross section and $\varepsilon_{wg} = -50$. In the 2D model system in Fig. 4a, the cylinder is made of metal and coated with a dielectric layer to guarantee a good dipole approximation. The permittivity of the metal is described by the Drude model $\varepsilon = 1 - \omega_p^2/(\omega^2 + i\omega\gamma)$ with $\omega_p = 3.57 \times 10^{15}$ rad s$^{-1}$ and $\gamma = 3 \times 10^{12}$ rad s$^{-1}$. The shell has a relative permittivity of 9. The cylinders have an inner radius of 5 nm and an outer radius of 6 nm. The waveguide has a relative permittivity of $\varepsilon_{wg} = -10$. The distance between the cylinders and the waveguide is $h_p = 25$ nm. The perturbation cylinder is 20 nm above the waveguide and has a permittivity of $-2 + i\delta$. $\delta$ controls the perturbation strength $\epsilon$. The value of $\epsilon$ was determined by calculating the reflection coefficient of the perturbation cylinder. All of the numerical simulations were performed using COMSOL Multiphysics[59].

**Microwave measurement.** The spheres have diameters of 5.56 mm and are made of $ZrO_2$ with a relative permittivity of $\varepsilon_d = 33.7 + 0.06i$. The waveguide has a cross section of $w \times t = 9.9 \text{ mm} \times 3.0 \text{ mm}$ and was cut from a commercial ECCOSTOCK dielectric slab with a relative permittivity of $\varepsilon_{wg} = 9 + 0.02i$. The spheres were separated from the waveguide by a 4-mm-thick foam spacer. The probe for detecting magnetic field has a split-ring shape with a radius of 1.5 mm. The detecting point is located 1 mm above the spheres. The ends of the waveguide were inserted into foam absorbers to realise the absorbing boundary condition.

**Data availability.** The authors declare that all data supporting the findings of this study are available within the paper and its supplementary information files. Additional data related to this paper are available from the corresponding authors upon reasonable request.

**Acknowledgements**

This work was supported by the Research Grants Council of the Hong Kong Special Administrative Region, China (No. AoE/P-02/12 and No. CityU 21302018) and the Major Program of Natural Science Research of Jiangsu Higher Education Institutions (No. 18KJA140003). S. Wang was also supported by a grant from City University of Hong Kong (No.

9610388). B. Hou acknowledges the visiting scholarship programme for youth scientists in Collaborative Innovation Center of Suzhou Nano Science and Technology and the Priority Academic Program Development (PAPD) of Jiangsu Higher Education Institutions. W. Lu acknowledges the support from Open Fund of the State Key Laboratory of Integrated Optoelectronics (IOSKL2017KF05).

**Author contributions**

S.W. conceived the idea and performed the calculations. B.H. designed and carried out the experiments. W.L. assisted the measurements. Y.T.C. and Z.Q.Z. assisted with the theoretical and numerical analysis. S.W. and C.T.C. supervised the project. All authors contributed to discussions and the writing of the manuscript.

**Additional information**

**Competing interests:** The authors declare no competing interests.

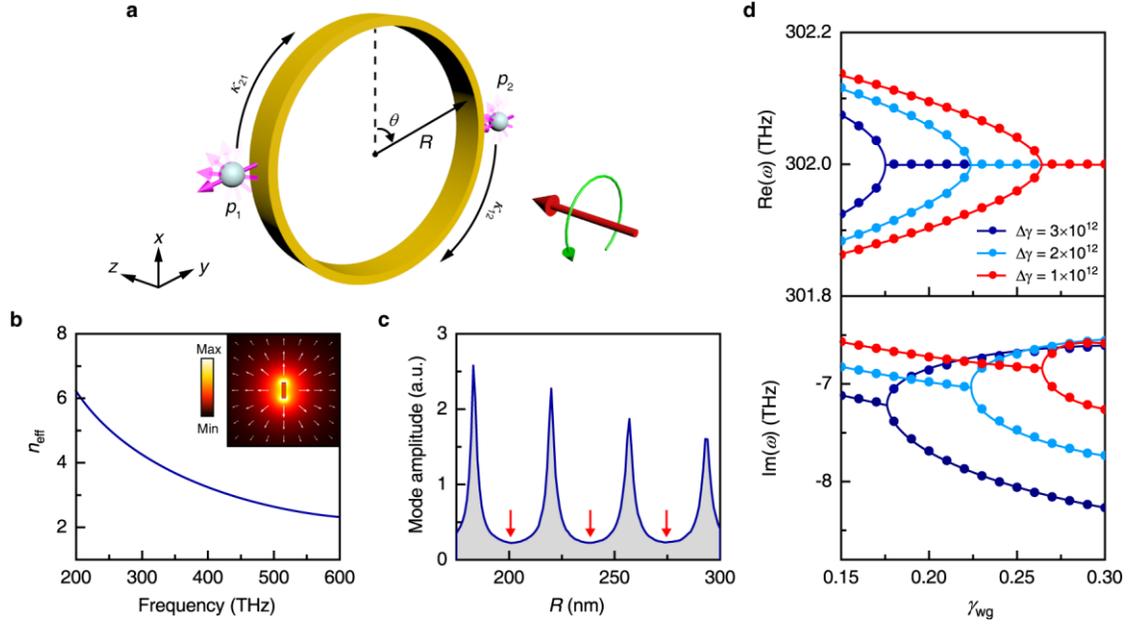

**Fig. 1** Exceptional point in coupled-dipole-resonator model system with mutual coupling. **a** Two spherical dipole resonators ($p_1$, $p_2$) placed near a circular plasmonic waveguide. Their chiral dipole resonance modes couple with each other through the unidirectionally propagating plasmon under spin-momentum locking. **b** Effective mode index of the guided plasmon. The insert shows the magnitude (colour) and direction (arrows) of the mode electric field. **c** Amplitude of the guided plasmon as a function of radius $R$ under the excitation of a point source near the circular waveguide. The peaks correspond to constructive interferences of the plasmon and the dips (marked by the red arrows) correspond to destructive interferences. **d** Real and imaginary parts of the eigenfrequencies as a function of the loss $\gamma_{wg}$ of the lower-half waveguide ($\theta \in [\pi/2, 3\pi/2]$) obtained by coupled mode theory. $\Delta\gamma$ is the loss difference of the two spheres.

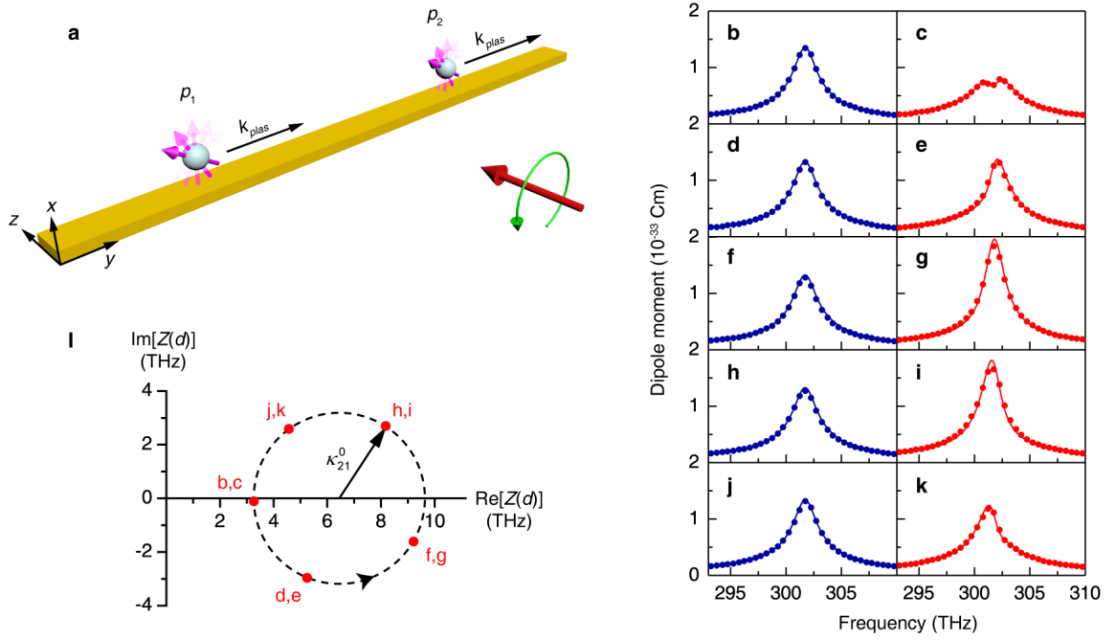

**Fig. 2** Exceptional point in the model system with unidirectional coupling. **a** In the limit of $\Delta\gamma \to 0$, the coupling from $p_2$ to $p_1$ is suppressed by the waveguide's loss. The model system reduces to the effective system where two resonators are unidirectionally coupled through a straight waveguide. **b–k** Evolution of the electric dipole moments $p_1$ (left column) and $p_2$ (right column) under left-handed circular polarisation excitation for $d = 2.0\lambda_{plas}$ (**b**, **c**), $2.2\lambda_{plas}$ (**d**, **e**), $2.4\lambda_{plas}$ (**f**, **g**), $2.6\lambda_{plas}$ (**h**, **i**) and $2.8\lambda_{plas}$ (**j**, **k**). The symbols denote the simulation results and the solid lines denote the fitting results of coupled mode theory. **l** Evolution of $Z(d) = \Gamma/2 - i\kappa_{21}(d)$ on the complex plane when $d$ varies. The red symbols (labelled by panel numbers) correspond to the fitting results in **b–k**.

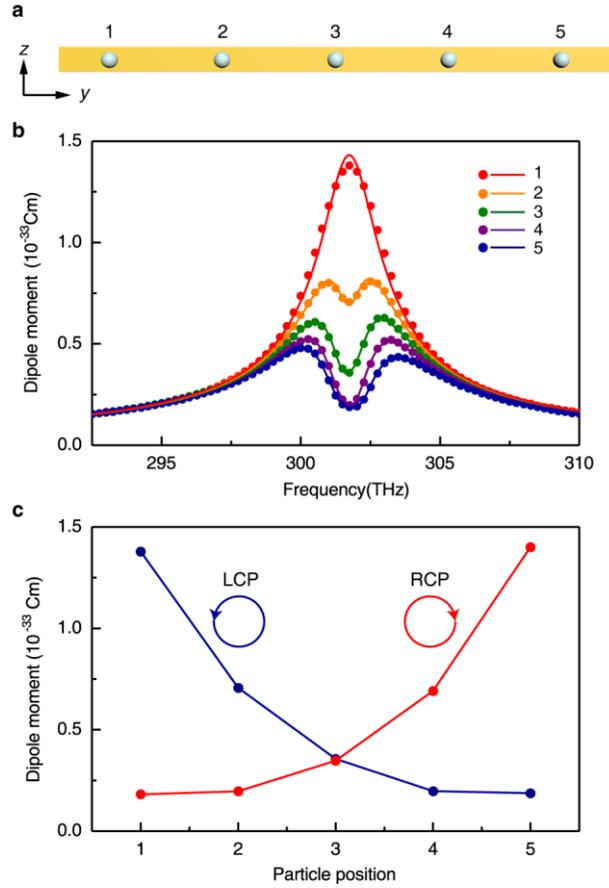

**Fig. 3** Higher-order exceptional point in the model system. **a** $N = 5$ spheres placed on the waveguide with equal spacing of $d = \lambda_{\text{plas}}$ (top view). **b** Electric dipole moments of the five spheres as a function of frequency under left-handed circular polarisation (LCP) excitation. The symbols denote the simulation results and the solid lines denote the results of coupled mode theory. **c** Resonant electric dipole moments of the five spheres as a function of position under LCP and right-handed circular polarisation (RCP) excitations.

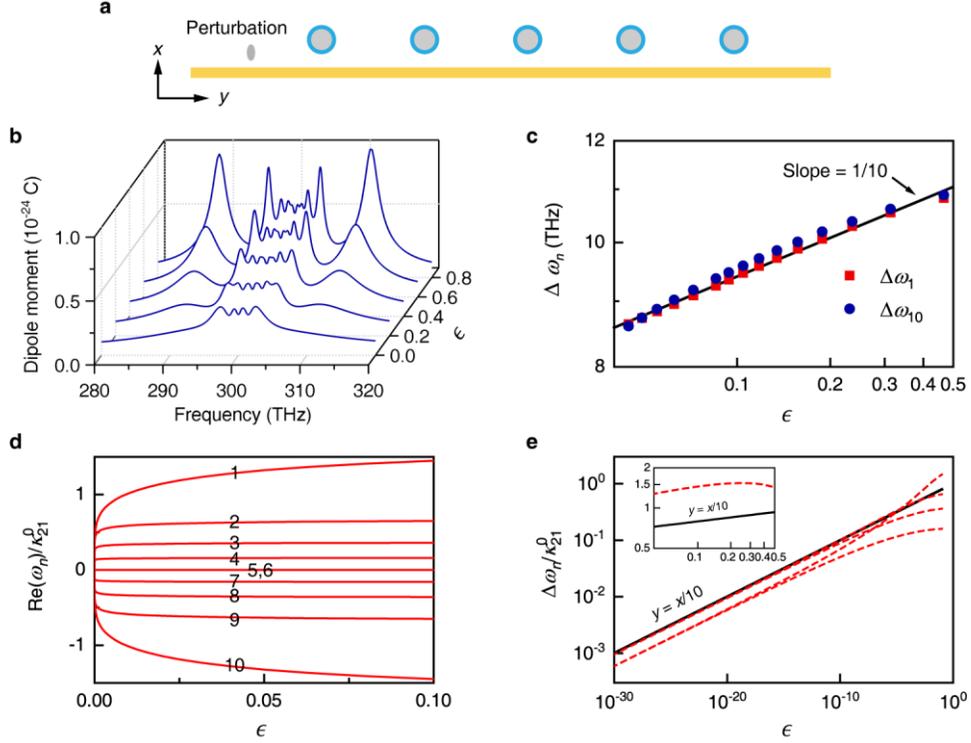

**Fig. 4** Enhancement of sensitivity by the higher-order exceptional point. **a** Two-dimensional model system consisting of five cylinders on a plasmonic waveguide. The spacing between cylinders is $5\lambda_{\text{plas}}$. A cylinder of elliptical shape is used to induce perturbation. **b** Total dipole moment of the cylinders as a function of frequency for different perturbation $\epsilon$. **c** Frequency splitting $\Delta\omega_n = |\omega_n - \omega_0|$ as a function of perturbation strength for $n = 1$ and $n = 10$, corresponding to the first and the last peaks in **b**. They follow a trend with a slope of 1/10 in the log–log plot. **d** Real parts of the normalised eigenvalues as a function of perturbation strength, obtained with the Hamiltonian in Eq. (5). **e** Normalised frequency splitting as a function of perturbation strength for the eigenvalues in **d**. When $\epsilon$ is small, all $\Delta\omega_n / \kappa_{21}^0$ follow the line of $y = x/10$. The inset shows a zoom-in of the large perturbation range for $\Delta\omega_1 / \kappa_{21}^0$.

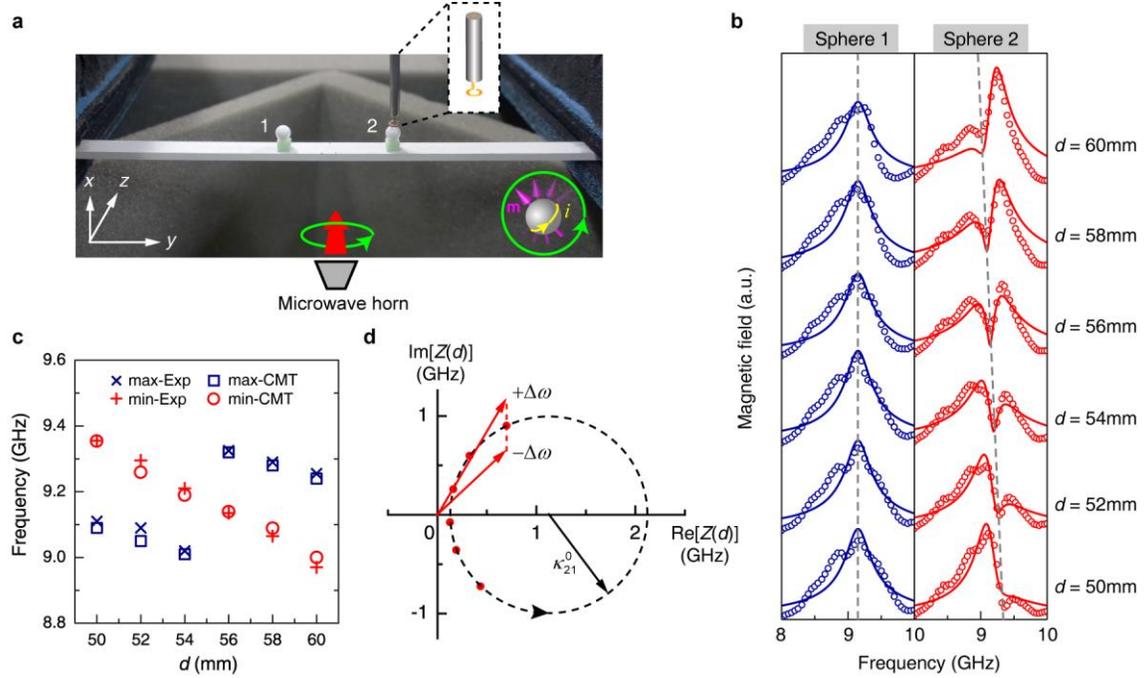

**Fig. 5** Microwave demonstration of the exceptional point. **a** Photograph of the experimental setup. Two identical dielectric spheres were placed on a dielectric strip waveguide with the gap distance controlled by a foam spacer. The spheres have diameters of 5.56 mm. A probe measured the magnetic field near the spheres under circular polarisation excitation. The inset in the right corner shows the chiral magnetic dipole mode. **b** Experimental results (symbols) and analytical fitting results (solid lines) of the magnetic field magnitude under left-handed circular polarisation excitation for different coupling distances $d$. The dashed lines mark the positions of the maxima (Sphere 1) and the minima (Sphere 2) of the field. **c** The position of the maxima and minima in the resonance spectra of Sphere 2 in panel **b**, showing a good match between the experimental (crosses) and analytical (circles and squares) results. **d** Evolution of $Z(d) = \Gamma/2 - i\kappa_{21}$ on the complex plane when $d$ varies. The red symbols correspond to the cases in panel **b**. The lengths of the red arrows denote the magnitude of $|\Gamma/2 - i\kappa_{21} \pm i\Delta\omega|$.

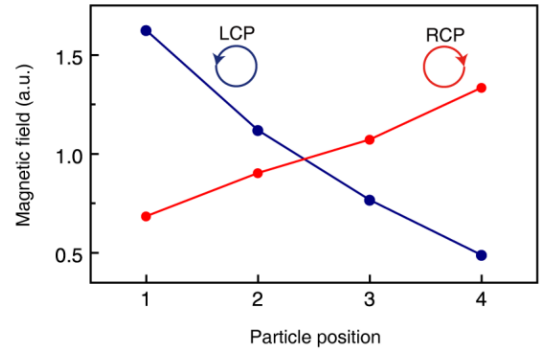

**Fig. 6** Measured magnetic field magnitude in the case of higher-order exceptional point. Four spheres with equal separations of $d = \lambda_{\text{plas}}$ were placed on the dielectric waveguide shown in Fig. 5a. The system is excited by a microwave of left-handed circular polarisation (LCP) or right-handed circular polarisation (RCP). The results show a monotonic trend that agrees with the model results in Fig. 3c.

Supplementary Information

# Arbitrary order exceptional point induced by photonic spin-orbit interaction in coupled resonators

Wang *et al*.

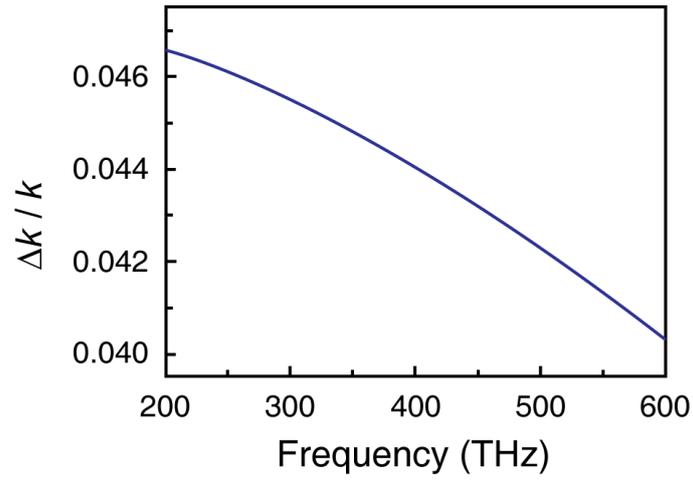

**Supplementary Figure 1 Variation of plasmon wavevector as a function of frequency for the lossy waveguide.** $k$ denotes the wavevector without loss and $\Delta k$ denotes the difference of the real part of the wavevector before and after introducing the loss into the waveguide. We have set $\gamma_{wg} = 0.3$. As $\Delta k / k$ is on the order of $10^{-2}$, the contribution of $\gamma_{wg}$ to the phase of the coupling parameters can be neglected.

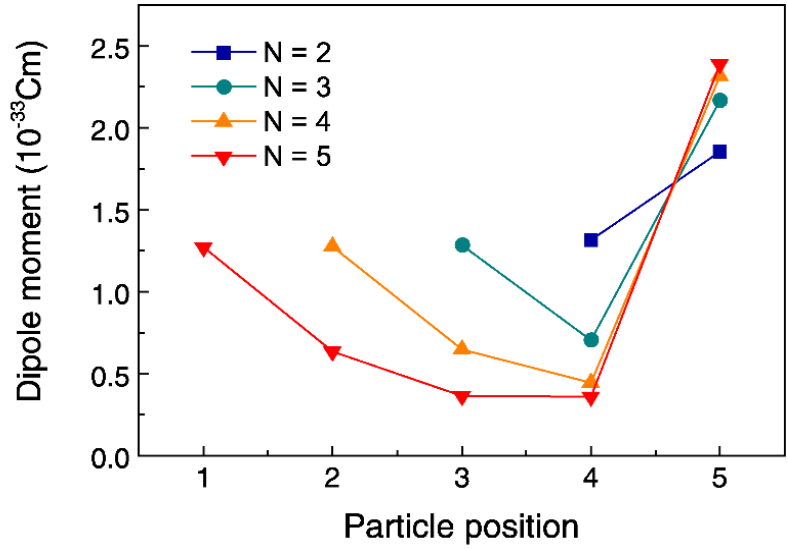

**Supplementary Figure 2 Higher order exceptional point for enhancing dipole emission.** Resonance electric dipole moment of each sphere under left-handed circular polarisation excitation for *N* spheres. Energy from the incident light is collected by the left spheres and transferred to the 5$^{th}$ sphere which has a larger dipole moment. The spacing of the 1-4$^{th}$ spheres is $d = \lambda_{plas}$ and the spacing between the 4th and the 5th spheres is $d = 1.5\lambda_{plas}$. With such a spacing arrangement, the guided wave coming from each of spheres 1-4 gives a constructive interference at the 5$^{th}$ sphere and hence an enhancement of dipole moment at that site. Note that the dipole moment at the last sphere increases as the number of spheres in the array increases.

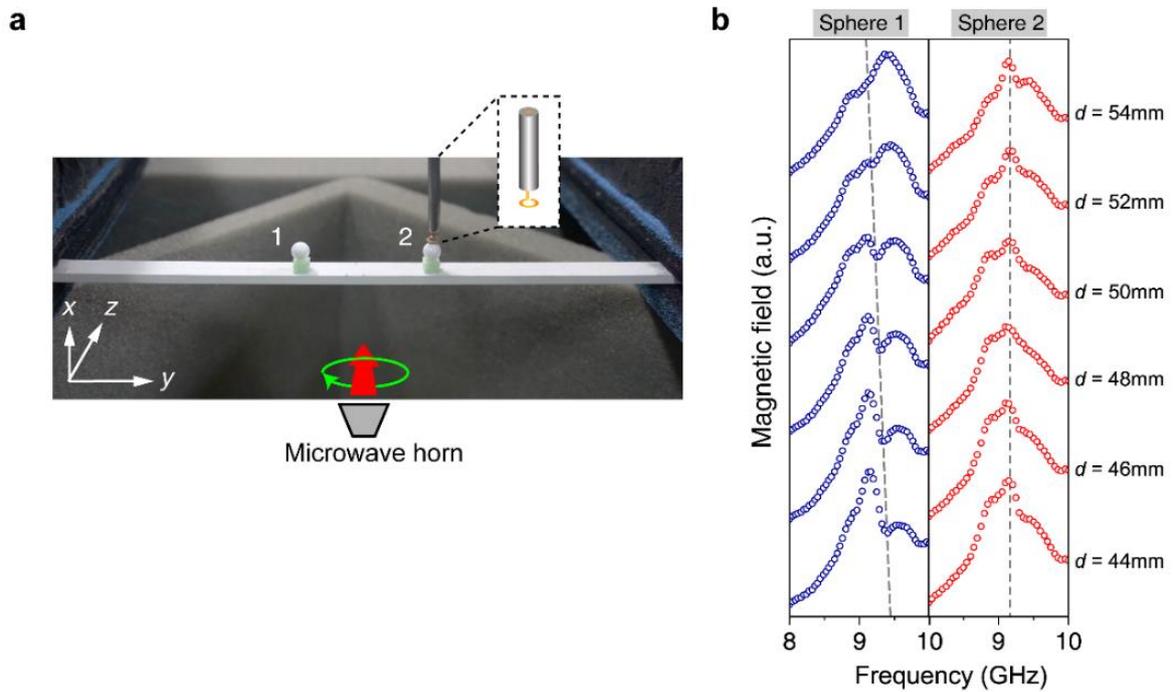

**Supplementary Figure 3 Measured magnetic dipole responses under right-handed circular polarisation excitation. a** Microwave experimental setup. Two identical dielectric spheres were placed on a dielectric strip waveguide with the gap distance controlled by a foam spacer. The spheres have diameters of 5.56 mm. A probe measured the magnetic field near the spheres under circular polarisation excitation. **b** Measured magnitude of the magnetic field on top of the two spheres. The dashed lines mark the positions of the local minima (Sphere 1) and the maxima (Sphere 2) of the field.

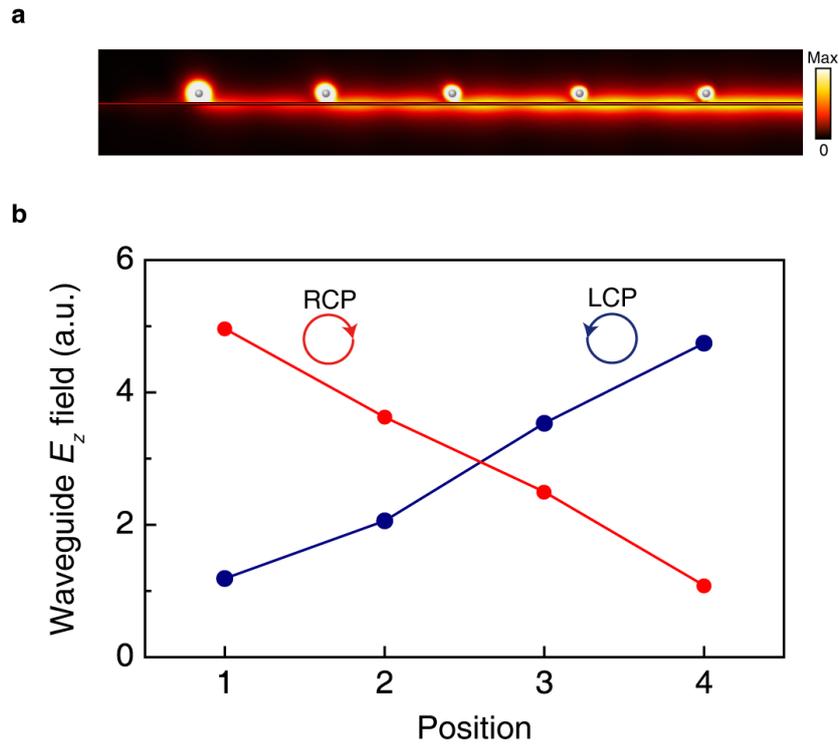

**Supplementary Figure 4 Position dependence of the waveguide field. a** Numerical simulation result of the electric field magnitude for the model system of Fig. 3a under left-handed circular polarisation excitation. **b** Measured $E_z$ field magnitude near the lower surface the waveguide (right below the spheres) for the microwave system of Fig. 6. Under left-handed (right-handed) circular polarisation excitation, the field increases from left (right) to right (left).

## Supplementary Note 1
## Monotonic behavior of induced dipole moments at higher-order exceptional points

Here we present an analytical derivation for the behaviors of chiral dipole moments at the EPs. Consider the case of LCP, the rate equation for $N$ coupled particles can be expressed as:

$$\frac{da_i}{dt} = -i\omega_0 a_i - \frac{\Gamma}{2} a_i - i\kappa_{ij} \sum_{j=1}^{i-1} a_j - \sqrt{\gamma_c} a_{in}, \tag{1}$$

Take the $N = 3$ case as an example, substitute the harmonic wave expressions $a_{in} = A_{in} e^{-i\omega t}$ and $a_i = A_i e^{-i\omega t}$ into the above equation, we obtain

$$(\omega - \tilde{\omega}_0) A_1 + i\sqrt{\gamma_c} A_{in} = 0, \tag{2}$$

$$(\omega - \tilde{\omega}_0) A_2 - \kappa_{21} A_1 + i\sqrt{\gamma_c} A_{in} = 0, \tag{3}$$

$$(\omega - \tilde{\omega}_0) A_3 - \kappa_{32} A_2 - \kappa_{31} A_1 + i\sqrt{\gamma_c} A_{in} = 0, \tag{4}$$

where $\tilde{\omega}_0 = \omega_0 - i\Gamma/2$. The static solution to the above equations is

$$A_1 = \frac{-i\sqrt{\gamma_c} A_{in}}{\omega - \tilde{\omega}_0}, \tag{5}$$

$$A_2 = \frac{-i\sqrt{\gamma_c} A_{in}}{(\omega - \tilde{\omega}_0)^2} (\omega - \tilde{\omega}_0 + \kappa_{21}), \tag{6}$$

$$A_3 = \frac{-i\sqrt{\gamma_c} A_{in}}{(\omega - \tilde{\omega}_0)^3} \left[ (\omega - \tilde{\omega}_0)^2 + (\omega - \tilde{\omega}_0)(\kappa_{32} + \kappa_{31}) + \kappa_{32}\kappa_{21} \right]. \tag{7}$$

At $\omega = \omega_0$, we have $\omega - \tilde{\omega}_0 = i\Gamma/2$. Assume the particle array has a period of $d = \lambda_{plas}$ as in Fig. 3a, we have $\kappa_{ij} = \kappa_{21} = -i\kappa_{21}^0, \forall i > j$, and the electric dipole moments can be expressed as

$$p_1 = \left| \frac{2\sqrt{\gamma_c} A_{in}}{\Gamma} \right|, \quad p_2 = \left| \frac{2\sqrt{\gamma_c} A_{in}}{\Gamma} \left( \frac{\Gamma - 2\kappa_{21}^0}{\Gamma} \right) \right|, \quad p_3 = \left| \frac{2\sqrt{\gamma_c} A_{in}}{\Gamma} \left( \frac{\Gamma - 2\kappa_{21}^0}{\Gamma} \right)^2 \right|. \tag{8}$$

Similar derivations can be done for arbitrary number of spheres, in which case the electric dipole moments are

$$p_i = \left| \frac{2\sqrt{\gamma_c} A_{in}}{\Gamma} \left( \frac{\Gamma - 2\kappa_{21}^0}{\Gamma} \right)^{i-1} \right|. \tag{9}$$

Both $\kappa_{12}^0$ and $\Gamma$ take positive real values and $|(\Gamma - 2\kappa_{21}^0)/\Gamma| < 1$ according to the results in Fig. 2l, so we have $p_i < p_{i-1}$, which indicates a monotonic decreasing trend of the dipole moments.